\documentclass[
final,
twocolumn,
letterpaper,
]{elsarticle}

\usepackage{graphicx}          
\usepackage{dcolumn}           
\usepackage{bm}                
\usepackage{hyperref}          
\usepackage[mathlines]{lineno} 
\usepackage{amsmath}
\usepackage{color}

\def\HOMO{^{\rm HOMO}}
\def\LUMO{^{\rm LUMO}}
\def\la{{\langle\, }} 
\def\ra{{\,\rangle }} 
\def\p{\,|\,} 
\def\TDKS{^{\rm TDKS}}
\def\br{{\bf r}} 
\def\GS{^{\rm GS}}
\def\hT{{\hat T}} 
\def\hS{{\hat S}} 
\def\hV{{\hat V}} 
\def\hH{{\hat H}} 
\def\hU{{\hat U}} 
\def\hW{{\hat W}} 
\def\hv{{\hat v}} 
\def\hp{{\hat p}} 
\def\hr{{\hat{\bf r}}} 
\def\hP{{\hat P}} 
\def\hR{{\hat{\bf R}}} 
\def\hj{{\hat j}} 
\def\hJ{{\hat J}} 
\def\hbr{{\hat{\bf r}}}

\def\hbJ{{\hat{\bf J}}} 
\def\hn{{\hat n}} 
\def\hN{{\hat N}} 
\def\bxt{{({\bf x}, t)}} 
\def\brt{{({\bf r}, t)}} 
\def\bRt{{({\bf R}, t)}} 

\def\bA{{\bf A}}

\def\sn{_{\rm n}}
\def\se{_{\rm e}}
\def\ee{_{\rm ee}} 
\def\en{_{\rm en}} 
\def\nn{_{\rm nn}} 
\def\exte{_{\rm ext,e}} 
\def\extn{_{\rm ext,n}} 
\def\Ne{{N_{\rm e}}} 
\def\Nn{{N_{\rm n}}} 
\def\rel{{\bf \underline{r} }} 
\def\rnucl{{\bf \underline{R} }} 
\def\br{\underline{\bf r}}
\def\bR{\underline{\bf R}}
\def\bx{\underline{\bf x}}

\graphicspath{
  {figures/pdf/}
  {figures/eps/}
}

\begin{document}


\title{Stochastic quantum molecular dynamics for finite and extended systems}

\author[fhi,ucsd,etsf]{Heiko Appel\corref{cor1}}
\address[fhi]{Fritz-Haber-Institut der Max-Planck-Gesellschaft,
              Faradayweg 4-6, D-14195 Berlin, Germany}
\ead{appel@fhi-berlin.mpg.de}
\cortext[cor1]{Corresponding author:}
\author[ucsd]{Massimiliano Di Ventra}
\address[ucsd]{University of California San Diego, La Jolla,
               California 92093, USA}
\address[etsf]{European Theoretical Spectroscopy Facility}
\date{\today}

\begin{abstract}
We present a detailed account of the technical aspects
of stochastic quantum molecular dynamics, an approach introduced
recently by the authors [H. Appel and M. Di Ventra, Phys. Rev. B
{\bf 80} 212303 (2009)] to describe coupled electron-ion dynamics
in open quantum systems. As example applications of the method we
consider both finite systems with and without ionic motion, as well
as describe its applicability to extended systems in the limit of
classical ions. The latter formulation allows the study of important
phenomena such as decoherence and energy relaxation in bulk systems
and surfaces in the presence of time-dependent fields.
\end{abstract}


\maketitle


\section{Introduction}

Time-dependent density-functional theory (TDDFT) calculations are currently
enjoying a large popularity due to their efficiency and
success in describing low-lying excitation energies in molecular
systems \cite{marques-2006}. In addition, many applications have been
investigated with TDDFT. Examples include
electronic transport \cite{diventra-2004,bushong-2005,burke-2005,kurth-2010}, 
nonlinear optical response \cite{vila-2010}, or atoms and molecules
in strong laser fields \cite{petersilka-1999,kreibich-2003}. In the latter
cases, the time-dependent Kohn-Sham (TDKS) equations are usually evolved
in real-time.
However, the majority of these studies pertains to the
description of closed quantum systems, since
the corresponding TDKS equations describe a set of N particles evolving
coherently in time.~\footnote{Notable exceptions are the
references~\cite{burke-2005,diventra-2007,dagosta-2008,zheng-2010,yuen-zhou-2009}.}
On the other hand, most experimental
situations involve some level of energy dissipation and/or decoherence
induced by either some environments to
which the given system is coupled, or the measurement apparatus itself
which necessarily projects non-unitarily the state of the system onto
states of the observables. This is generally true for both electrons and
ions, so that a first-principles description of their coupled dynamics in
the presence of one or more environments is of fundamental importance in order
to describe phenomena and compare with experiments. At this point, it
is worth noting that present quantum molecular dynamics (QMD) approaches,
(e.g., the Born-Oppenheimer, Ehrenfest or Car-Parrinello methods) either
do not allow excited states dynamics (Born-Oppenheimer and Car-Parrinello
methods) or, if they do (e.g., Ehrenfest QMD), they do not permit electronic
coupling to external environments. Indeed, in all these approaches, energy
dissipation and thermal coupling to the environment are usually described
with additional thermostats coupled directly to the classical nuclear
degrees of freedom, which fall short of describing the numerous physical
phenomena associated with decoherence and energy dissipation.

In order to overcome these shortcomings, we have recently introduced a novel
time-dependent density functional approach based on stochastic time-dependent
Kohn-Sham equations \cite{appel-2009}, where we allow the coupling of both
electrons and (in principle quantum) ions with external baths. This approach -
we have named stochastic quantum molecular dynamics (SQMD) - extends the
previously introduced stochastic time-dependent-current density-functional
theory (STDCDFT)~\cite{diventra-2007,dagosta-2008} to the coupled dynamics of
electrons and ions. The latter was formulated to account for electrons
interacting with external environments, without however including atomic
motion. Therefore, SQMD combines and improves on the strengths of STDCDFT
and present QMD methods by greatly expanding the physical range of applications
of these methods.

Clearly, from a practical point of view the present approach
suffers - like all density-functional theory (DFT) based methods - from
our limited knowledge of the properties of the exact exchange-correlation
functional. Furthermore, in the present case, the exact functional depends
not only on the electronic degrees of freedom, but also on the ionic and
bath(s) degrees of freedom \cite{appel-2009}. Nevertheless, due to the weak
system-bath(s) coupling assumption of the present theory, as well as the
limited number of systems where quantum nuclear effects are of disproportionate
importance, we may start by considering the limit of SQMD to classical nuclei
and adopt the available functionals of standard closed-system TDDFT.
Like in any other practical application of DFT, it is the predictions that we
obtain and comparison with experiments that will be the ultimate judge of the
range of validity of the approximate functionals used.

In Ref.~\cite{appel-2009} we have outlined the details of the proof of the
theorem at the core of SQMD, and provided a simple example of the
relaxation dynamics of a finite system (a molecule) prepared in some excited
state and embedded in a thermal bath. However, there are some technical details
behind an actual implementation of this approach we have not reported yet,
and which are nonetheless important if one is interested in using this method
for practical computations. In this work we then present all the technical
aspects for a practical implementation and use of SQMD. In addition, we
present the theory behind its applicability to extended systems which is of
great importance in the study of decoherence and energy relaxation in bulk
systems and surfaces. We are in the process of implementing SQMD for extended
systems and we will report these results in a forthcoming
publication \cite{appel-2011}.

The paper is organized as follows. In Section \ref{sec:theory} we give an
introduction to the theory of stochastic quantum molecular dynamics. For
completeness, this includes general aspects of open quantum systems as well
as the basic theorem of SQMD. In Section \ref{sec:simulation_algorithms} we
discuss the aspects of a practical implementation of SQMD.
Finally, in Section \ref{sec:applications} we illustrate with
some examples the application of SQMD to finite systems with and
without ionic motion, and outline its extension to periodic systems.
Conclusions are reported in Section \ref{sec:conclusions}.

\section{Theory}
\label{sec:theory}

\subsection{Stochastic Schr\"odinger equation}
In the following, we consider an electron-ion many-body system
coupled to a bosonic bath. For simplicity, we will
consider only a single bath, but the formulation is trivially extended to the
case of several environments. The total Hamiltonian of the entire system
is then

\begin{equation}
\label{eq:TotalHamiltonian}
\hat H = \hat H_S\otimes \hat I_B + \hat I_S\otimes \hat H_B
         + \lambda \hat H_{SB}.
\end{equation}
Our system of interest is described by the many-body Hamiltonian
$\hat H_S$ and the environment degrees of freedom are given in terms
of $\hat H_B$.
The interaction of the system with the environment is given by
the Hamiltonian $\hat H_{SB}$ and is assumed to be
weak in the sense that a perturbation expansion in terms of this coupling
can be performed. With $\lambda$ we denote the corresponding coupling
parameter for the system-bath interaction.

The total system described by the Hamiltonian $\hat H$ follows
a unitary time-evolution, which can be formulated for pure states
either in terms of the time-dependent Schr\"odinger equation (TDSE), with $\hbar=1$
\begin{equation}
\label{eq:TDSE}
i \partial_t \Psi(t) = \hat H(t) \Psi(t),
\end{equation}
or, alternatively for mixed states, in terms of the Liouville-von
Neumann equation
\begin{equation}
\label{eq:LvN}
\frac{d}{dt}\hat\rho(t) = -i\left[ \hat H(t), \hat\rho(t)\right],
\end{equation}
where $\hat\rho$ is the statistical operator.

Since $\hat H$ is a many-body Hamiltonian and we have to deal, in
principle, with infinitely many degrees of freedom (due to the bath)
it is not possible to solve Eq.~(\ref{eq:TDSE}) or (\ref{eq:LvN}) in
practice, except for a few simple model cases. In
addition, in most cases of interest the microscopic knowledge
about the bath is limited and only its macroscopic thermodynamic properties
are known, e.g., one typically assumes that the bath is in
thermal equilibrium. However, we are only
interested in the dynamics of the system degrees of freedom. It is
therefore desirable to find an effective description for the system only.

To accomplish this we may trace out the bath degrees of
freedom at the level of the statistical operator, namely we perform the operation
$\hat\rho_S=\textrm{Tr}_B\{\hat\rho\}$, where $\hat\rho_S$ is called the reduced
statistical operator of system $S$. It is worth pointing out here that this
procedure does not generally lead to a closed equation of motion for the
reduced statistical operator and one needs further approximations. Depending on
the approximations involved, one may arrive at an
effective quantum master equation for the reduced density operator
$\hat\rho_S$ \cite{vankampen-2007,breuer-2002,weiss-2008}. As we will discuss later,
this approach has some drawbacks when used within a density-functional formulation,
both fundamental - in view of the theorems of DFT - and practical, since
solving for the density matrix is computationally more demanding than
solving directly for state vectors.

We therefore take here a different route. Instead of working
with a derived/composite quantity like the statistical operator,
we summarize briefly how the bath degrees of freedom can be
traced out directly at the level of the
wavefunction. The derivation that follows has been reported elsewhere
in the literature (see, e.g., Ref. \cite{gaspard-1999}). We repeat some
steps here for completeness and to clarify our starting point.

To this end let us consider the set of eigenfunctions
$\{\chi_n(x_B)\}$ of the bath Hamiltonian
\begin{equation}
\label{eq:BathEigenvalueProblem}
\hat H_B \chi_n(x_B) = \varepsilon_n \chi_n(x_B),
\end{equation}
with $x_B$ the bath's coordinates (including possibly spin),
and expand the total wavefunction of Eq.~(\ref{eq:TDSE})
in the complete set of orthonormal states formed
by $\{\chi_n(x_B)\}$, namely~\footnote{
At first sight this expansion might seem formally similar to the
factorization used for the Born-Oppenheimer (BO) approximation.
However, in the BO case the expansion
coefficients depend on the dynamical variables of both subsystems,
electrons and nuclei. This dependence originates from the fact that the
electronic Hamiltonian in the Born-Oppenheimer approximation has
a parametric dependence on the nuclear degrees of freedom.
In the present case we assume that the partitioning of the
total Hamiltonian in Eq.~(\ref{eq:TotalHamiltonian}) is such that
expansion (\ref{eq:TotalWavefunctionExpansion}) becomes exact.
}

\begin{equation}
\label{eq:TotalWavefunctionExpansion}
\Psi(x_S,x_B;t) = \sum_n \phi_n(x_S;t)\chi_n(x_B),
\end{equation}
with $\phi_n(x_S;t)$ some functions (not necessarily normalized)
in the Hilbert space of the system $S$.

In order to see that in the presence of a bath the functions
$\phi_n(x_S;t)$ form a statistical ensemble describing the
properties of the subsystem $S$, let us proceed as follows.
For a general observable $\hat O_S$ of the system $S$ we find
after simple algebra (and using the orthonormality of the
bath states $\chi_n(x_B)$)
\begin{equation}
\begin{split}
\langle \Psi(x_S,x_B;t)|\hat O_S &|\Psi(x_S,x_B;t)\rangle = \\
& \sum_n \langle \phi_n(x_S;t)|\hat O_S| \phi_n(x_S;t) \rangle.\label{averageAS}
\end{split}
\end{equation}
Let us now normalize the functions $\phi_n(x_S;t)$ by writing
\begin{equation}
\psi_n(x_S,t)\equiv \phi_n(x_S;t)/p_n(t)
\end{equation}
with
\begin{equation}
p_n(t)=\langle \phi_n(x_S;t)|\phi_n(x_S;t)\rangle,
\end{equation}
which, according to Eq.~(\ref{eq:TotalWavefunctionExpansion}),
is nothing other than the probability for the bath to be in the
state $\chi_n(x_B)$.

We can now define the following statistical operator
\begin{equation}
\begin{split}
\hat \rho_S&\equiv \sum p_n(t)|\psi_n(x_S;t)\rangle \langle \psi_n(x_S;t)|\\
&\equiv \overline{|\psi(t)\rangle \langle \psi(t)|},
\end{split}
\end{equation}
and immediately recognize that the average~(\ref{averageAS}) can be
re-written as
\begin{equation}
\langle \Psi(x_S,x_B;t)|\hat O_S |\Psi(x_S,x_B;t)\rangle =
\textrm{Tr}\{ \hat \rho_S O_S\},\label{averageAS1}
\end{equation}
namely, due to the interaction with the bath, the system $S$ is
necessarily in a mixture of states defined by the macrostate
$\{p_n(t),\psi_n(x_S;t)\}$. We thus expect that the equation of
motion for the representative wave-function $\psi(t)$ of the
subsystem $S$ to be ``naturally'' stochastic, namely
we expect to find an equation of motion that provides the
macrostate $\{p_n(t),\psi_n(x_S;t)\}$.

In order to show this we follow the Feshbach projection-operator method
\cite{feshbach-1958,nordholm-1975} and define the following projection
operators
\begin{align}
\label{eq:ProjectionOperators}
\hat P_n &:= \hat I_S \otimes \p \chi_n \ra \la \chi_n \p, \\
\hat Q_n &:= \hat I_S \otimes \sum_{k\ne n} \p \chi_k \ra \la \chi_k \p,
\end{align}
where $\hat I_S$ is the identity in the Hilbert space of the system.
The rationale behind the choice of the above operators is to obtain the
equation of motion of a representative coefficient $\phi_n(x_S;t)$.

By acting with these projection operators on the many-body TDSE for
the combined system and bath in Eq.~(\ref{eq:TDSE}) we arrive at
\begin{equation}
i\partial_t \hat P_n \Psi(t) =
     \hat P_n \hat H \hat P_n \Psi(t) + \hat P_n \hat H \hat Q_n \Psi(t)
\label{Pproj}
\end{equation}
\vspace{-4mm}
\begin{equation}
i\partial_t \hat Q_n \Psi(t) =
     \hat Q_n \hat H \hat Q_n \Psi(t) + \hat Q_n \hat H \hat P_n \Psi(t)
\label{Qproj}
\end{equation}
Equation~(\ref{Qproj}) can be formally solved. Inserting the result
back into Eq.~(\ref{Pproj}) we obtain
\begin{equation}
\label{eq:EffectiveSystemWavefunction}
\begin{split}
i\partial_t & \hat P \Psi(t) = \hat P \hat H \hat P \hat P \Psi(t)
  + \hat P \hat H \hat Q e^{-i \hat Q\hat H\hat Q t} \hat Q \Psi(0) \\
  & -i\int_0^t \,\hat P \hat H \hat Q e^{i \hat Q\hat H\hat Q (\tau-t)}
        \hat Q \hat H \hat P \hat P \Psi(\tau)\,d\tau,
\end{split}
\end{equation}
where we have omitted the index $n$ for brevity.
The first term on the right hand side of Eq.~(\ref{eq:EffectiveSystemWavefunction})
contains only projections on the system manifold, and describes the coherent
evolution of the system degrees of freedom. The second term is a source term that
carries a dependence on initial conditions ($\hat Q \Psi(0)$ are the initial
conditions of all system's states except the one we are considering), and
the third term on the right hand side is a memory term that is recording
the history of the time evolution.\footnote{These equations have a formal
similarity to the quantum transport formulation introduced by Kurth et.~al. 
\cite{kurth-2005}.
However, in this case the projection operators project on the real space regions
of leads and central molecular device. Also the bath is fermionic in the quantum
transport case (leads) and electrons can be exchanged between system and ``bath''.
This is in contrast to the present case where we consider bosonic baths
and only energy and momentum can be exchanged between system and bath.}

Note that, up to this point, we have made no approximations,
i.e., the time evolution given by
Eq.~(\ref{eq:EffectiveSystemWavefunction}) is still fully coherent.
However, the solution of Eq.~(\ref{eq:EffectiveSystemWavefunction}) is
very involved and, apart from model systems, not feasible in practice.
Furthermore, a solution would require the initial conditions for all
the microscopic degrees of freedom of the bath. These cannot all be
determined simultaneously by a measurement. In practice, one rather
has only knowledge about macroscopic thermodynamic properties of the
bath, like temperature and pressure.
It is therefore common to perform the following additional
approximations which are motivated by the form of the system-bath
interaction and the thermodynamic properties of the bath:
(i) due to the assumed weak coupling between system and bath the source 
and memory terms are expanded up to second
order in the system-bath coupling parameter $\lambda$,
(ii) the bath and subsystem $S$ are assumed to be uncorrelated at the
initial time, (iii) a random phase approximation is
performed for the phases in the
source and memory terms\footnote{The random phase approximation
invoked in the derivation of the Markovian stochastic Schr\"odinger
equation might seem at first sight surprising. The derivation of
the Lindblad equation from the Markovian stochastic Schr\"odinger
equation on the other hand shows, that both describe the exact
same dynamics if the Hamiltonian does not depend on internal
degrees of freedom or any time-dependent or stochastic field
(see Sec.~\ref{Lindblad}).},
and (iv) it is assumed that the bath degrees of
freedom form a dense energy spectrum and are in local thermal equilibrium
characterized by
\begin{equation}
\hat \rho_B = \frac{1}{\mbox{Tr}(e^{-\beta \hat H_B })}\,e^{-\beta \hat H_B},
\end{equation}
where $\beta = 1/ {k_B T}$.

Let us then write the interaction Hamiltonian as
\begin{equation}
\hat H_{SB}=\sum_{\alpha} \hat S_{\alpha}  \otimes \hat B_{\alpha},
\end{equation}
where $\hat S_{\alpha}$ and $\hat B_{\alpha}$ are - in the most general case - many-body
operators that act on the Hilbert spaces of the system and bath, respectively.
In the following we will also assume that the average of the operators $\hat B_{\alpha}$
vanishes on the n-th eigenstate of the bath, namely
\begin{equation}
\sum_{\alpha} \hat S_{\alpha} \la \chi_n \p \hat B_{\alpha}\p \chi_n \ra=0.
\end{equation}
If this is not the case we simply redefine the system Hamiltonian via
\begin{equation}
\hat H'_S=\hat H_S+\lambda \sum_{\alpha} \hat S_{\alpha} \la \chi_n \p \hat B_{\alpha}\p \chi_n \ra,
\end{equation}
and the interaction Hamiltonian
as $\hat H'_{SB}=\hat H_{SB}-\lambda \sum_{\alpha} \hat S_{\alpha} \la \chi_n \p \hat B_{\alpha}\p \chi_n \ra$.
The term $\la \chi_n \p \hat B_{\alpha}\p \chi_n \ra$ thus contributes to the unitary
evolution of the system by renormalizing its eigenvalues (a typical example of
this is the Lamb shift \cite{breuer-2002,weiss-2008}).

With these approximations in place, the
source term can be regarded as a stochastic driving term. This is because,
the system's state we have singled out in
Eq.~(\ref{eq:ProjectionOperators}) now interacts with a (practically infinite)
large set of bath states densely distributed in energy. The previously
coherent equation Eq.~(\ref{eq:EffectiveSystemWavefunction}) then has to be regarded
as a non-Markovian stochastic Schr\"odinger equation for the general state
vector $\psi(t)\equiv \phi_n(x_S;t)/\langle \phi_n(x_S;t)|\phi_n(x_S;t)\rangle$
\cite{gaspard-1999}
\begin{equation}
\label{eq:NonMarkovianStochasticTDSE}
\begin{split}
i \partial_t & \psi(t) = \hat H_S \psi(t)
   + \lambda \sum_\alpha l_\alpha(t) \hat S_\alpha \psi(t) \\[-2mm]
  & -i\lambda^2 \sum_{\alpha\beta} \int_0^t C_{\alpha\beta}(t-\tau)
    \hat S^\dagger_\alpha e^{-i \hat H_S(t-\tau)} \hat S_\beta \psi(\tau) d\tau \\
 &+ {\cal O}(\lambda^3),
\end{split}
\end{equation}
where $l_\alpha(t)$ are stochastic processes with zero ensemble average,
$\overline{l_\alpha(t)}=0$, and correlation functions
\begin{equation}
\overline{l_\alpha(t)l_\beta(t')} = 0,
\end{equation}
\begin{equation}
\overline{l^*_\alpha(t)l_\beta(t')} = C_{\alpha\beta}(t-t').
\end{equation}
Equation~(\ref{eq:NonMarkovianStochasticTDSE}) is a general non-Markovian
stochastic Schr\"odinger equation. Indeed, it still contains a time-integral
over the past dynamics which is originating from the memory term
of Eq.~(\ref{eq:EffectiveSystemWavefunction}). Even though the theorem of
SQMD could be formulated with non-Markovian baths we will focus in the
following only on the Markovian limit
\begin{equation}
\label{eq:MarkovianApproximation}
C_{\alpha\beta}(t-t') \propto \delta_{\alpha\beta}\delta(t-t'),
\end{equation}
namely, we consider baths that are
$\delta$-correlated. Physically, this means that the bath does not retain memory
of the interaction with the system which is valid when the typical thermalization
time-scales inside the bath are much faster than the thermalization time-scales
of the system. This approximation is well justified for a large number of
bath degrees of freedom. If this assumption does not hold, one has to resort to
the solution of the more involved Eq.~(\ref{eq:NonMarkovianStochasticTDSE}).

By inserting the Markov approximation,
Eq.~(\ref{eq:MarkovianApproximation}), into
Eq.~(\ref{eq:NonMarkovianStochasticTDSE}) we then arrive at the
stochastic Schr\"odinger equation in the Born-Markov limit
\begin{equation}
\begin{split}
\label{eq:MarkovianTDSE}
i \partial_t  \psi(t) = & \hat H_S(t) \psi(t)
     - \frac{i}{2}\sum_{\alpha}
     \hat S^{\dagger}_\alpha \hat S_\alpha \psi(t)\\
    & + \sum_\alpha l_\alpha(t) \hat S_\alpha \psi(t),
\end{split}
\end{equation}
where the parameter $\lambda$ has been absorbed in the
operators $\hat S_\alpha$. The first term on the right hand side of
Eq.~(\ref{eq:MarkovianTDSE}) is the usual unitary evolution of the
system under the action of the system Hamiltonian $\hat H_S$, the second term
describes the dissipation effects introduced by the bath and
would indeed make the probability density generated by $\psi(t)$
decay in time. The last term, however, introduces fluctuations so that
the norm of the state vector $\psi(t)$ averaged over the ensemble is
conserved, namely
$\langle \overline {\psi(t)|\psi(t)}\rangle=1+{\cal O}(\lambda^4)$.

Due to the stochastic nature of this equation, the stochastic
process described by Eq.~(\ref{eq:MarkovianTDSE}) has to be
simulated in terms of an ensemble of state vectors $\psi(t)$.
Each member $\psi(t)$ of the ensemble evolves differently in
time due to the random variables $l_\alpha(t) $ in the
third term on the rhs. of Eq.~(\ref{eq:MarkovianTDSE}). If we consider an
initial mixed state
\begin{equation}
\label{eq:ReducedStatisticalOperatorin}
\hat \rho_S(0) = \sum_k p^0_k \p \psi_k(0) \ra \la \psi_k(0)\p,
\end{equation}
where $p^0_k$ are the probabilities (with $\sum_k p^0_k=1$) of
finding the state $\psi_k(0)$ in the ensemble, the statistical
average over all members of the ensemble
allows us to construct the reduced density operator for
the system degrees of freedom
\begin{equation}
\label{eq:ReducedStatisticalOperator}
\hat \rho_S(t) = \sum_k p^0_k \overline{ \p \psi_k(t) \ra \la \psi_k(t)\p}.
\end{equation}
Here, we use the symbol $\overline{\cdots}$ to indicate the
statistical average over all members of the ensemble of state
vectors $\psi_k(t)$, namely the ensemble $\{\psi^i_k(t)\}$ of state
vectors with initial conditions $\psi_k(0)$.

The expectation value of a general physical observable of the system $S$, $\hat O_S$,
can then be computed as in Eq.~(\ref{averageAS1}), i.e.,
\begin{equation}
\begin{split}
\la \hat O_S \ra
 & = \mathrm{Tr}(\hat O_S \hat \rho_S(t)) \\
 & = \mathrm{Tr}(\hat O_S \sum_k p^0_k \overline{ \p \psi_k(t) \ra \la \psi_k(t)\p} )\\
 & = \sum_k p^0_k \overline{\la \psi_k(t) \p \hat O_S \p \psi_k(t) \ra},
\end{split}
\end{equation}
where the last step shows that the construction of $\hat \rho_S(t)$ is
not actually required: we can compute expectation values of observables
directly from the wave-functions in the usual way, followed by a
statistical average over all members of the ensemble of state vectors.
It is also important to note that this approach provides {\it directly}
the {\it full} distribution of the given observable at any given time,
provided we can compute a large enough set of realizations of the
stochastic processes $l_\alpha(t)$ (for an example of this see, e.g.,
Ref. \cite{appel-2009}). From this distribution we can then compute all
higher moments and/or cumulants (e.g., the variance, skewness, etc.) some of which
are directly accessible experimentally.

\subsection{Derivation of the Lindblad equation and stochastic Hamiltonians}
\label{Lindblad}
For many-body Hamiltonians which are not {\it stochastic}, namely they do not depend on
internal degrees of freedom - Hamiltonians
$\hat H_S\neq \hat H_S[\{|\psi^j_k\rangle\}]$ - or do not depend explicitly
on some stochastic field, like e.g., a stochastic thermostat \cite{bussi-2007}, it is possible to derive the
Lindblad equation \cite{lindblad-1976,gardiner-1985,breuer-2002,weiss-2008}
from the stochastic Schr{\"o}dinger equation (\ref{eq:MarkovianTDSE}).

For notational clarity, let us denote in the following
discussion with $|\psi\rangle$ a {\em single member} of the stochastic
ensemble $\{|\psi^j_k\rangle\}$.
If we consider for simplicity the case of a single bath
operator in Eq.~(\ref{eq:MarkovianTDSE}), and observe that
in the Markovian limit
\begin{equation}
W(t)=\int_0^tl(t') dt'
\label{intdW}
\end{equation}
is a Wiener process \cite{vankampen-2007}
with properties $\overline{dW} = 0$ and $\overline{dW^\dagger dW}=dt$, we can
formulate the stochastic Schr{\"o}dinger equation (\ref{eq:MarkovianTDSE})
for a single bath in differential form according to
\begin{equation}
d|\psi\rangle=\left[-i\hat H_S |\psi\rangle -\frac12 \hat S^\dagger
\hat S |\psi\rangle\right]dt -i \hat S |\psi \rangle dW.
\label{eq:stochasticse-fin}
\end{equation}

Next, we employ It\^o stochastic calculus in order to compute the
following differential
\begin{equation}
d |\psi\rangle\langle\psi| = (d |\psi\rangle) \langle\psi| +
|\psi\rangle (d \langle\psi|) + (d |\psi\rangle) (d \langle\psi|).
\label{eq:differential_product}
\end{equation}
Unlike in normal calculus, we also have to keep the third term
in the product rule above. This becomes necessary, since a
statistical average over the Wiener increment $dW^\dagger dW$ is proportional
to $dt$, which will cause terms quadratic in $dW$ to contribute
to {\em first} order in $dt$. Inserting Eq. (\ref{eq:stochasticse-fin})
and its Hermitian conjugate into Eq. (\ref{eq:differential_product})
we arrive after elementary algebra at
\begin{align}
\label{eq:lindblad_differential}
d |\psi\rangle\langle\psi| = \nonumber
  & -i \hat S |\psi\rangle\langle\psi| dW + \mbox{h.c.} \nonumber \\
  & -i \Big[ \hat H_S, |\psi\rangle\langle\psi| \Big] dt
    - \frac{1}{2} \Big\{ \hat S^\dagger \hat S, |\psi\rangle\langle\psi| \Big\} dt \nonumber \\
  & + \hat S |\psi\rangle\langle\psi| \hat S^\dagger dW^\dagger dW \nonumber \\
  & + \hat S |\psi\rangle\langle\psi| \hat H_S\, dWdt + \mbox{h.c.} \\
  & + \frac{i}{2} \hat S |\psi\rangle\langle\psi|\hat S^\dagger \hat S\, dWdt
    + \mbox{h.c.} \nonumber \\
  & + \hat H_S|\psi\rangle\langle\psi| \hat H_S dt^2
    + \frac{1}{4} \hat S^\dagger \hat S |\psi\rangle\langle\psi| \hat S^\dagger \hat S dt^2 \nonumber \\
  &  + \frac{i}{2} \Big\{ \hat H_S , |\psi\rangle\langle\psi|\hat S^\dagger \hat S \Big\} dt^2. \nonumber
\end{align}

In order to construct the statistical operator from the state vectors
of the statistical ensemble $\{|\psi^j_k\rangle\}$,
we perform in the next step the statistical average over all members
in the ensemble, i.e.
\begin{equation}
d \hat\rho = d \overline{|\psi\rangle\langle\psi|}.
\end{equation}
Taking the properties $\overline{dW} = 0$, $\overline{dWdt} = 0$
and $\overline{dW^\dagger dW}=dt$
of the stochastic process $l(t)$ into account, we see that only the second
and third line in Eq. (\ref{eq:lindblad_differential}) contribute to
first order in $dt$ and we arrive at
\begin{align}
\label{eq:average_dm}
d \hat\rho =
 & -i \overline{\Big[ \hat H_S, |\psi\rangle\langle\psi| \Big]} dt
   - \frac{1}{2}  \Big\{ \hat S^\dagger \hat S, \overline{|\psi\rangle\langle\psi|} \Big\} dt \nonumber \\
 & + \hat S \overline{|\psi\rangle\langle\psi|} \hat S^\dagger dt + O(dt^{2}).
\end{align}
At this point, note that this equation of motion is not necessarily
closed for $\hat\rho =  \overline{|\psi\rangle\langle\psi|}$ because
the first term on the right hand side of
Eq.~(\ref{eq:average_dm}) is not equal to the commutator
$-i \left[\hat H_S,\hat \rho_S\right]$ {\it unless}
$\hat H_S \neq \hat H_S[\{|\psi^j_k\rangle\}]$, or $\hat H_S$ 
does not depend on any stochastic field, or the system is in
a pure state at all times - which would amount to the case 
$\hat S=0$.\footnote{A further complication would
arise if the operators $\hat S$ depended on internal degrees of 
freedom, i.e., $\hat S=\hat S[\{|\psi^j_k\rangle\}]$.
In that case, the average over the statistical ensemble in the 
second and third terms on the right hand side of 
Eq.~(\ref{eq:average_dm}) has to be
performed over the operators $\hat S$ as well.}
However, if the Hamiltonian is stochastic,
one has to deal with
an {\it ensemble of Hamiltonians}, and the statistical average of
the first term on the right hand side of Eq.~(\ref{eq:average_dm})
involves also a statistical average over these Hamiltonians (see, e.g.,
Refs. \cite{balian-1991,diventra-2008}).

For the moment being, let us assume that
$\hat H_S\neq \hat H_S[\{|\psi^j_k\rangle\}]$ and furthermore that
the Hamiltonian $\hat H_S$ does not depend on some external stochastic field.
In this case we find
\begin{equation}
\partial_t\hat \rho_S = -i \left[\hat H_S, \hat \rho_S \right] -\frac12
\left\{ \hat S^\dagger \hat S, \hat \rho_S \right\} + \hat S \hat \rho_S\hat S^\dagger
\label{density-matrix}
\end{equation}
which is the well-known quantum master equation in Born-Markov limit
(or Lindblad equation if the bath operators and the Hamiltonian, do
not depend on time)~\cite{lindblad-1976,gardiner-1985,breuer-2002,weiss-2008}.

We have thus shown that the stochastic Schr\"odinger equation
of Eq.~(\ref{eq:MarkovianTDSE}) and the master equation
(\ref{density-matrix}) lead to the same statistical operator,
if and only if the Hamiltonian is not stochastic. However, in 
order to prove any DFT theorem one is led to consider the 
dynamics of the actual many-body system and that of any 
{\it auxiliary} one (including the Kohn-Sham system) with 
different interaction potentials, but reproducing the exact 
many-body density or current density. It is then at this stage 
that a choice has to be made - in the case of a many-body 
system open to one or more environments - regarding the basic 
equation of motion to work with. If we choose to work with a 
quantum master equation of the type (\ref{density-matrix}), 
then we are {\it assuming} from the outset that the Kohn-Sham 
Hamiltonian is not stochastic. But this is an hypothesis that 
constitutes part of the final thesis, namely we have to prove 
that this statement is correct, not assume it {\it a priori} 
\cite{diventra-2009}. This issue does not arise with the stochastic
Schr{\"o}dinger equation (\ref{eq:MarkovianTDSE}), because in 
that case we can consider all possible Hamiltonians, including 
those that are stochastic.

In addition to the above important point, we also recall that 
for arbitrary time-dependent operators $\hat S(t)$ and 
$\hat H_S(t)$, Eq.~(\ref{density-matrix}) may not yield a 
positive-definite statistical operator at all times (see, e.g.,
Ref.~\cite{maniscalco-2004}). This is a major limitation in 
practical calculations, since loss of positivity (which 
precludes a statistical interpretation of physical observables) 
should then be checked at every instant of time. Note that such 
a limitation does not pertain to the stochastic Schr\"odinger 
equation which can be equally applied to arbitrary time-dependent 
operators without possible loss of positivity of the ensuing 
statistical operator. Therefore, the above two issues make the 
equation of motion of the statistical operator a less solid 
starting point for a DFT theory of open quantum systems.

\subsection{Theorem of Stochastic QMD}
We are now in a position to state the basic theorem of SQMD.
Before doing this let us define the basic quantities we work with.
The many-body system we are interested in consists of $\Ne$ electrons with
coordinates $\br \equiv \{ {\bf r}_j \}$ and $N_n = \sum_s N_{s,n}$
nuclei, where each nuclear species $s$ comprises $N_{s,n}$ particles with
charges $Z_{s,j}$, masses $M_{s,j}$, $j=1\ldots N_{s,n}$, and
coordinates $\bR \equiv \{ {\bf R}_{s,j} \}$, respectively. 
Their dynamics - subject to an arbitrary classical electromagnetic 
field, whose vector potential is $\bA(t)$ - is described by the Hamiltonian
\begin{align}
\label{eq:TotalHamiltonianWithNuclei} \hH_S(t) & =
     \hT\se(\rel, t)   + \hW\ee(\rel)
   + \hU\exte(\rel, t) + \hW\en(\rel,\rnucl) \nonumber \\
  & + \hT\sn(\rnucl, t) + \hW\nn(\rnucl) + \hU\extn(\rnucl, t),
\end{align}
where $\hT\se(t)$ and $\hT\sn(t)$ are the kinetic energies of
electrons and ions, with velocities $\hv_k(t) = [ \hp_k +
e\bA(\hr_k, t) ]/m$ and $\hV_\alpha(t) = [ \hP_\alpha - Z_\alpha
\bA(\hR_\alpha, t) ]/M_\alpha$, respectively
and $\hU\exte(\rel, t)$, $\hU\extn(\rnucl, t)$ the
external potentials acting on electrons and ions. The particle-particle
interactions are given by
\begin{equation}
\begin{split}
\hW\ee(\rel) &= \frac{1}{4\pi\epsilon_0}\sum_{{j < k}}^\Ne
  \frac{e^2}{|\hr_j - \hr_k|}
\equiv  \sum_{{j < k}}^\Ne w\ee(\hr_j - \hr_k),\\
\hW\nn(\rnucl) &= \frac{1}{4\pi\epsilon_0}\sum_{ \alpha < \beta }^\Nn
  \frac{Z_\alpha\,Z_\beta}{|\hR_\alpha - \hR_\beta|} \\
& \equiv  \sum_{{\alpha < \beta}}^\Nn w\nn(\hR_\alpha - \hR_\beta),\\
\hW\en(\rel,\rnucl) &=
 - \frac{1}{4\pi\epsilon_0}\sum_{k=1}^\Ne
   \sum_{\alpha=1}^\Nn \frac{e\,Z_\alpha}{|\hr_k - \hR_\alpha|} \\
 & \equiv \sum_{k=1}^\Ne \sum_{\alpha=1}^\Nn w\en(\hr_k -
 \hR_\alpha).
\end{split}
\end{equation}
We then define the charge current operator
\begin{equation}
\hj\brt = \frac{e}{2m} \sum_k \{\hv_k(t),\delta(\bf{r} - \hbr_k) \}
\end{equation}
for the electrons, and
\begin{equation}
\hJ_\alpha\bRt =
   \frac{Z_\alpha}{2M_\alpha}\hspace{-5mm}\sum_{\stackrel{\beta}
       {Z_\alpha = Z_\beta,\, M_\alpha = M_\beta}}\hspace{-5mm}
\{\hV_\beta(t), \delta(\bf{R} - {\hat{\bf R}}_\beta) \},
\end{equation}
for the ions.
The total particle and current density
operators of the system can then be written as
\begin{equation}
\hN\bxt = \hn\brt +
\sum_\alpha \hN_\alpha\bRt
\end{equation}
for the particle number, and
\begin{equation}
\hbJ\bxt = \hj\brt + \sum_\alpha
\hJ_\alpha\bRt
\end{equation}
for the current density. To
simplify the notation we have also denoted with $\bx \equiv \{ \bR, \br \}$ the
combined set of electronic and nuclear coordinates, and we use the
combined index $\alpha = \{s,j\}$ for the nuclear species.

We now formulate the theorem for a single bath operator. It trivially extends
to many operators.
{\em Theorem.---} For a given bath operator $\hat S$, many-body
initial state $\Psi(\bx, t=0)$ (not necessarily pure) and external
vector potential $\bA(\bx,t)$, the dynamics of the stochastic
Schr\"odinger equation in Eq.~(\ref{eq:MarkovianTDSE}) generates
ensemble-averaged total particle and current densities $\overline{N(\bx, t)}$
and $\overline{{\bf J}(\bx,t)}$. Under reasonable physical assumptions,
any other vector potential $\bA'(\bx,t)$ (but same initial state
and bath operator) that leads to the same ensemble-averaged total
particle and current density, has to coincide, up to a gauge
transformation, with $\bA(\bx,t)$.

A sketch of the proof of this theorem can be found in the original
paper~\cite{appel-2009}. We thus refer the reader to this publication
for more details. Here, we just mention an important point. As in
Ref.~\cite{diventra-2007,dagosta-2008} we are implicitly assuming
that given an initial condition, bath operator, and ensemble-averaged
current density, a unique ensemble-averaged density can be obtained
from its equation of motion:
\begin{align}
\label{eq:contopenA}
\frac{\partial \overline {N(\bx, t)}}{\partial t} = &
-\nabla\cdot\overline {{\bf J}(\bx,t)} \\
&+\left \langle \overline{\hat
S^\dagger \hat n \hat S - \frac12 \hat S^\dagger \hat S \hat n -
\frac12 \hat n \hat S^\dagger \hat S } \right
\rangle. \nonumber
\end{align}
If we write this equation in the compact form
\begin{equation}
\partial_t \overline {N(\bx, t)}=-\nabla\cdot \overline {{\bf J}(\bx,
t)}+\mathcal{F}_B(\bx,t)
\label{eq:contopenA1}
\end{equation}
the above amounts to saying that $\mathcal{F}_B(\bx,t)$ is a
functional of $\overline{N(\bx, t)}$ and $\overline{{\bf J}(\bx,t)}$,
or better of $\overline{{\bf J}(\bx,t)}$ alone, and that
Eq.~(\ref{eq:contopenA}) admits a unique physical solution. Therefore,
unlike what has been recently argued~\cite{yuen-zhou-2009}, the
density is not independent of the current density, and our theorem
establishes a one-to-one correspondence between current density and
vector potential. If this were not the case, namely that the particle
and current densities were {\it independent} functions, then
$\mathcal{F}_B(\bx,t)$ would not be completely determined by the sole
knowledge of $\overline{N(\bx, t)}$ and $\overline{{\bf J}(\bx,t)}$ \cite{diventra-2009}.

\subsection{The limit of classical nuclei and Kohn-Sham scheme}
At this stage we may formulate a Kohn-Sham
(KS) scheme of SQMD where an exchange-correlation (xc) vector
potential ${\bf A}_{xc}$ - functional of the initial states, bath
operator(s), and ensemble-averaged current density - acting on
non-interacting species, allows to reproduce the exact
ensemble-averaged density and current densities of the original
interacting many-body system. The ensuing charge and current
densities would contain {\it all} possible correlations in the system
- if the exact functional were known.

However, in the present case, we
could construct several schemes - based on corresponding theorems -
by defining different densities
and current densities. For instance, we could collect all nuclear densities into
one quantity as done in Ref.~\cite{kreibich-2001}.
This, by no means is a limitation of this approach. Rather, it allows us to
``specialize" the given schemes to
specific physical problems. Instead, a much more serious limitation relates to the
construction of xc functionals for the chosen scheme.
Therefore, as anticipated in the introduction, we will restrict ourselves
here to the limit of classical nuclei.

Let us then assume that we know the vector potential
$\bA_{\rm eff}$ that generates the exact current density in the
non-interacting system. By construction, the system follows the
dynamics induced by the stochastic Schr\"odinger equation (for a single bath operator)
\begin{equation}
\begin{split}
d|\Psi_{KS}\rangle =&\left(-i\hat H_{KS}-\frac12 \hat S^\dagger \hat S\right)|\Psi_{KS}\rangle dt \\
 &-i\hat S |\Psi_{KS}\rangle dW
\label{ksse}
\end{split}
\end{equation}
where $|\Psi_{KS}\rangle$ is a Slater determinant of single-particle
wave-functions and
\begin{equation}
\hat H_{KS}=\sum_{k=1}^N \frac{\left[
{\bf {\hat p}}_k+e{\bf A}_{\rm eff}(\hbr_k,\rnucl,t)\right]^2}{2m} \label{ksh}
\end{equation}
is the Hamiltonian of non-interacting particles with
\begin{equation}
{\bf A}_{\rm eff}(\hbr_k,\rnucl,t)={\bf A}_{\rm ext}(\hbr_k,\rnucl,t) +{\bf A}_{\rm hxc}(\hbr_k,\rnucl,t),
\end{equation}
where ${\bf A}_{\rm ext}$ is the vector potential applied to the {\em true}
many-body system, and ${\bf A}_{\rm hxc}$ is the vector potential
whose only scope is to mimic the correct dynamics of the
ensemble-averaged current density, and we have lumped in
it also the Hartree interaction potential in addition to the
xc one. All these potentials depend on the instantaneous
classical nuclear coordinates $\rnucl$.

We immediately note that for a general many-body bath operator acting on many-body
states one cannot reduce Eq.~(\ref{ksse}) to a set of
independent single-particle equations, as done in the usual DFT
schemes for closed systems. In other words, this would generally require the
solution of an equation of motion of Slater determinants, which is still
computationally quite demanding. To see this point, suppose we have $N$ particles and retain $M$
single-particle states. We then need to solve for $C_N^M-1$
elements of the state vector (with $C_N^M=M!/N!(M-N)!$ and the $-1$
comes from the normalization condition). In addition, one has to
average over an amount, call it $m$, of different realizations of
the stochastic process. The problem thus scales exponentially
with the number of particles. If this seems prohibitive let
us also recall that a density-matrix formalism would
be even more computationally demanding, requiring the solution of
$(C_N^M +2)\times (C_N^M-1)/2$ coupled differential equations, even
after taking into account the constraints of hermiticity and unit
trace of the density matrix.

It was recently suggested in Ref.~\cite{pershin-2008}
that for operators of the type $\hat O=\sum_j \hat O_j$,
namely operators that can be written as sum over
single-particle operators (like the density or current
density), the expectation value of $\hat O$ over a many-particle
non-interacting state with dissipation can be approximated as a sum
of single-particle expectation values of $\hat O_j$ over an ensemble
of $N$ single-particle systems with specific single-particle
dissipation operators. In particular, it was found that the approximate single-particle
scheme provides an excellent approximation for the current
density compared to the
exact many-body calculation.~\cite{pershin-2008} We refer the reader to
Ref.~\cite{pershin-2008} for the numerical demonstration of
this scheme and its analytical justification.

From now on, for numerical convenience, we will then adopt the same {\em ansatz}
which in the present case reads,
\begin{equation}
\overline{\langle \Psi_{KS}|\hat O| \Psi_{KS}\rangle} \simeq
\sum_{j=1}^\Ne \overline{\langle \phi^j_{KS}|\hat O_j|
\phi^j_{KS}\rangle},\label{ansatzsp}
\end{equation}
with $|\phi^j_{KS}\rangle$ single-particle KS states solutions of
\begin{align}
\label{eq:kssesp}
d|\phi^j_{KS}\rangle = & -i \frac{\left( {\bf \hat
p}+e{\bf A}_{\rm eff}(\hbr,\rnucl,t)\right)^2}{2m} |\phi^j_{KS}\rangle dt \\
& -\frac12 \hat S^{j\dagger}_{sp} \hat
S^{j}_{sp}|\phi^j_{KS}\rangle dt
-i\hat S^{j}_{sp} |\phi^j_{KS}\rangle dW(t), \nonumber
\end{align}
with $\hat S^{j}_{sp}$ an operator acting on single particle states.\\

\subsection{Model for the system-bath interaction}
\label{sec:system_bath_interaction_model}

The single-particle operators $S^j_{sp}$ in Eq. (\ref{eq:kssesp}) that we employ 
in the present work are given by
the following time-independent projectors \footnote{Cf. also to the examples
in Refs.~\cite{pershin-2008,bushong-2008,dagosta-2008,appel-2009}.}
\begin{align}
\label{eq:bath_operators}
\hat S^j_{kk'}(\br) =\,
 & \delta_{kj} (1 - \delta_{kk'}) \sqrt{\gamma_{jk'}(\br)
               f_D(\epsilon_k) } \nonumber \\
 & \times \p \psi_j^{\rm GS}(\br) \ra \la \psi_{k'}^{\rm GS}(\br)  \p,
\end{align}
where
$f_D({\bf \varepsilon}_k) = \left[ 1+\exp\left(\frac{\varepsilon_k - \mu}{k_B T}\right) \right]^{-1}$
denotes the usual Fermi-Dirac distribution and $ \delta_{kj}(1 - \delta_{kk'})$
denote Pauli blocking factors.
The projectors in Eq. (\ref{eq:bath_operators}) cause a relaxation
of the system back to the ground-state orbitals $\p \psi_j^{\rm GS}(\br) \ra$
with a rate given by the rate constants $\gamma_{jk}(\br)$ 
(generally space and orbital dependent), while the
temperature in the Fermi factor
is modeling the temperature of the bath.

For the present paper, we further assume that the
rates are independent of space and orbital indexes,
i.e, $\gamma_{jk}(\br) \propto 1/\tau$, where $\tau$
is a relaxation time.
The bath operators of this model are sufficient for the illustration
purposes of the present work but they clearly provide only a simplified
picture of the full system-bath interaction.
We emphasize here, that a rigorous form of the bath operators and the
associated relaxation rates can always be derived from the microscopic
form of the complete Hamiltonian of system and bath and their
mutual interaction, Eq.~(\ref{eq:TotalHamiltonian}). For example,
in the case of a phonon bath, the system-bath interaction Hamiltonian
could be taken into account in terms of e.g., a Fr\"ohlich interaction.
In that case the relaxation rates can be extracted from the electron-phonon
coupling matrix elements. The situation is much simpler is the case of a photon bath,
where the Einstein rates of stimulated and spontaneous emmission can
be used.

\subsection{Forces on ions}
Once we have the single-particle KS states and the corresponding Slater
determinant $\Psi_{KS}(\bx,t)$ at hand we can compute the forces on
the nuclei as \cite{appel-2009}
\begin{equation}
{\bf F}_{\alpha}(t)=-\langle
\Psi_{KS}(\bx,t)|\nabla_{\alpha} \hat H_{KS}(\bx,t)| \Psi
_{KS}(\bx,t)\rangle\,, \label{force}
\end{equation} for each realization of the stochastic
process.~\footnote{Note that Eq.~(\ref{force}) is {\it not} the
expression for the force one would obtain from the Hellmann-Feynman
theorem. This is because we are considering a system out of
equilibrium. Rather, Eq.~(\ref{force}) is the total time derivative
of the average of the ion momentum operator over the state of
the system (see, e.g., Ref. \cite{pantelides-2000}).}
Note, that this force is stochastic in nature since the wavefunctions
in the above expectation value are solutions of a stochastic Schr\"odinger
equation. Since approximations to the xc functional of the Kohn-Sham
Hamiltonian may make the latter stochastic then the force one would
obtain using a density matrix approach - e.g., by solving the quantum
master equation~(\ref{density-matrix}) - would not be necessarily equal
to the average force obtained from Eq.~(\ref{force}) by averaging
over the ensemble of realizations.

\section{Simulation Algorithms}
\label{sec:simulation_algorithms}

We have now outlined the general theory behind SQMD and we are ready
to move on to the description of its actual implementation.

\subsection{Real-time propagation}

The standard real-time propagation of the Kohn-Sham orbitals
for a closed quantum system is based on numerical approximations
for the time-ordered evolution operator
\begin{equation}
\hat U(t+\Delta t,t) = \hat T \exp\left( -i \int_t^{t+\Delta t}
                         \hat H_{\rm KS}(\tau) d\tau \right).
\label{eq:time_evolution_operator}
\end{equation}
There are several approaches employed in standard TDDFT computer packages,
like, e.g., {\tt octopus} \cite{castro-2006} to evaluate Eq.~(\ref{eq:time_evolution_operator})
numerically. Here, we have chosen the Magnus
propagator as basic building block for our stochastic simulations.~\cite{magnus-1954}
The Magnus series \cite{magnus-1954} provides an exact expression for the
time-evolution operator
(\ref{eq:time_evolution_operator}) as a {\em time-unordered} exponential of so
called Magnus operators $\Omega_j$ in the form
\begin{equation}
\hat U(t+\Delta t,t) = \exp\left( \hat \Omega_1 + \hat \Omega_2 + \hat \Omega_3 + \cdots \right),
\end{equation}
where the $\hat \Omega_j$ are given in terms of time-integrals over nested
commutators of the Hamiltonian at different points in time
\begin{equation}
\begin{split}
\hat \Omega_1 =& -i  \int_t^{t+\Delta t} \hat H_{\rm KS}(\tau) d\tau \\
\hat \Omega_2 =& \int_t^{t+\Delta t}\int_t^{\tau_1} [\hat H_{\rm KS}(\tau_1),
                     \hat H_{\rm KS}(\tau_2)] d\tau_2 d\tau_1 \\
&\vdots
\end{split}
\end{equation}
The time-integrals can be evaluated numerically with e.g., a Gauss-Legendre
quadrature. In the simplest case, which is accurate up to second
order in the time-step, one arrives at the exponential midpoint rule
\begin{equation}
\begin{split}
\hat U(t+\Delta t,t) &= \exp\left( \hat \Omega_1 \right) + O(\Delta t^3) \\
\hat \Omega_1 & = -i \hat H_{\rm KS}(t+\Delta t/2) + O(\Delta t^3).
\end{split}
\end{equation}
Higher-order approximations can be easily derived from the Magnus series
and appropriate quadrature points and weights, but experience shows that
the second order gives a good balance between speed and accuracy for many
applications. In the present work we use this approximation for the piecewise
deterministic evolution that we are going to introduce in the next section.

\begin{figure*}[ht]
 \includegraphics[width=\textwidth]{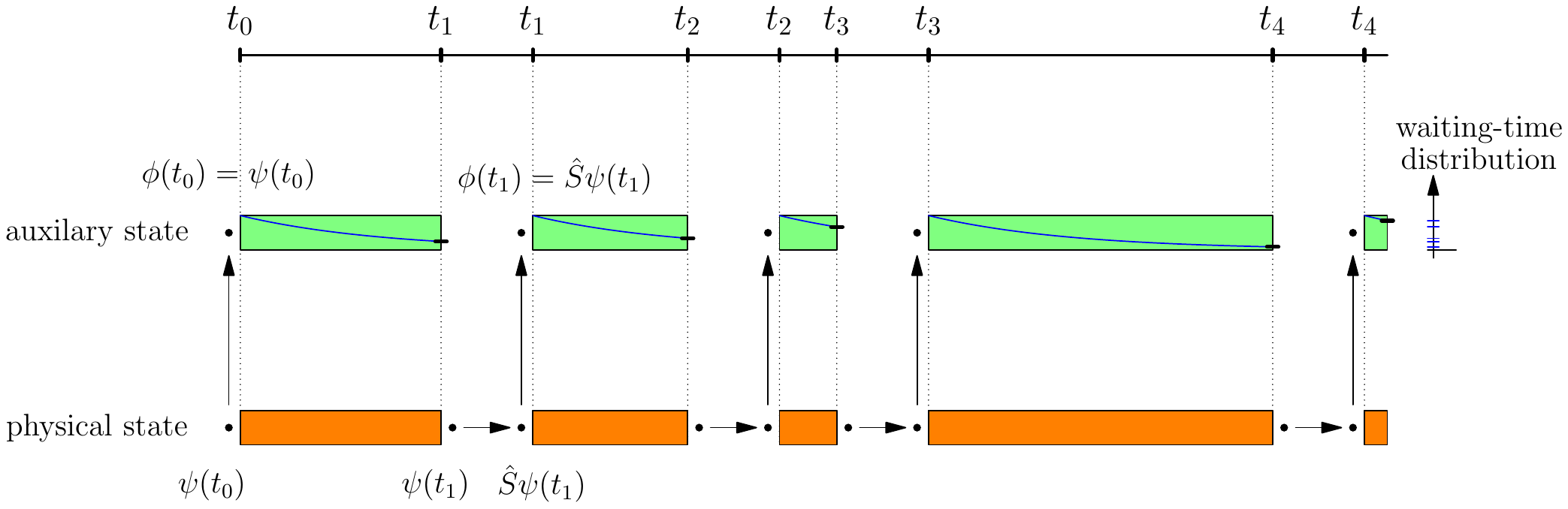}
 \caption{
  The figure illustrates the time-evolution as generated by the
  quantum jump algorithm. The lower track
  represents the piecewise deterministic propagation of the physical state
  which is intercepted at instances in time where the bath operator $\hat S$
  acts on the state. The points in time where this takes place are
  determined by sampling a waiting-time distribution. The sampling is
  performed by propagating an auxiliary state (represented in the upper track)
  with a non-Hermitian Hamiltonian. Uniformly distributed random numbers
  are drawn and once the norm of the auxiliary state drops below the current
  random number the propagation of the physical and the auxiliary state is
  suspended. At this point in time the action of the bath operator on the
  physical state results in a new state which is then also used to initialize
  the auxiliary state for the evolution. The simulation of both states is then
  resumed again.
 \label{fig:quantum_jump_algorithm}}
\end{figure*}

\subsection{Quantum Jump Algorithm}\label{QJA}

We are now left with the actual solution of the stochastic
Schr{\"o}dinger equation (\ref{eq:MarkovianTDSE}). In past work,
this has been done by directly integrating this equation with
standard approaches - e.g., with appropriately modified Runge-Kutta
methods (see e.g.~\cite{dagosta-2008,kloeden-1992} and references therein).
These approaches are reasonable when we deal with a small number
of accessible states or short propagation times. However, they
become increasingly unstable with an increasing number of states or
for very long timescales, which is the case for realistic systems,
like molecular structures, surfaces or solids.
As an alternative, we have thus adopted the quantum jump algorithm
pioneered in the work of Di\'osi \cite{diosi-1988}, Dalibard \cite{dalibard-1992},
Zoller and Gardiner and collaborators \cite{dum-1992a,gardiner-1992,dum-1992}
as well as Carmichael \cite{carmichael-1989}.
At the price of introducing the propagation of auxiliary states, the quantum jump
algorithm provides improved stability for systems with a large number of
states/particles and, due to the piecewise deterministic evolution,
also a stable propagation scheme for long timescales.

This algorithm works as follows.
Consider the deterministic time-evolution given by the follwing
norm-preserving non-linear Schr\"odinger equation
\begin{equation}
\label{eq:DeterministicTimeEvolution}
\frac{d}{dt}\psi_j(t) = -i \left(\hat {\cal H}_{S} + \frac{i}{2}
                || \hat S \psi ||^2 \right) \psi_j(t),
\end{equation}
where the non-Hermitian Hamiltonian ${\cal H}_S$ is given by
\begin{equation}
\hat {\cal H}_S = \hat H_S - \frac{i}{2} S^\dagger S.
\end{equation}
As before, the operators $\hat H_S$ and $ \hat S$ denote the 
Hermitian system Hamiltonian and the bath operator of 
Eq.~(\ref{eq:MarkovianTDSE}) \footnote{For convenience we 
consider here the case of a single bath. A generalization to many 
baths is straightforward.}.
The main objective of the quantum jump algorithm is to sample
the stochastic process given by Eq.~(\ref{eq:MarkovianTDSE}) in terms
of a piecewise deterministic evolution, i.e. a set of deterministic
time intervals generated by the evolution of 
Eq.~(\ref{eq:DeterministicTimeEvolution})
and action of the bath operator $\hat S$ between two consecutive time
intervals. A central ingredient of the algorithm is a waiting-time
distribution which determines when the jumps (i.e., actions
of the bath operator) appear throughout the simulation.

In order to sample the unknown waiting-time distribution, an auxiliary
set of wavefunctions $\phi_j^{\rm aux}$ is introduced. The wavefunctions 
$\phi_j^{\rm aux}$ are propagated with the non-Hermitian
Hamiltonian $\hat {\cal H}_S$ alongside the actual states $\psi_j$.
Since the auxiliary system evolves with a non-Hermitian Hamiltonian,
the norm of the states $\phi_j^{\rm aux}$ is not preserved. It can
be shown \cite{breuer-2002} that the decay of the norm of
the auxilary wavefunctions is related to the waiting-time distribution. 
The algorithm makes use of this fact and directly samples the 
waiting-time distribution on the fly from the norm decay. 

In terms of the Kohn-Sham system, 
the steps of the algorithm can then be summarized as follows
\begin{itemize}
\item[1)] Draw a uniform random number $\eta_k \in [0,1]$ for the
          Kohn-Sham Slater determinant
\item[2)] Propagate $N$ auxiliary orbitals $\phi_j^{\rm aux}$ under the non-Hermitian dynamics
\begin{equation}
\label{phieq}
i\partial_t \phi_j^{\rm aux} = \left[ \hH_{\rm KS} - \frac{i}{2}\hS^\dagger\hS\right]
 \phi_j^{\rm aux}, \qquad j=1\ldots N \nonumber
\end{equation}
\item[3)] Propagate the orbitals $\psi_j^{\rm KS}$, $j=1\ldots N$  of the Kohn-Sham
          system with a norm-conserving dynamics according to
\begin{equation}
i\partial_t \psi_j^{\rm KS} = \left[ \hH_{\rm KS}
  - \frac{i}{2}\hS^\dagger\hS
  + i ||\hat S\psi_j^{\rm KS}||^2 \right] \psi_j^{\rm KS} \nonumber
\end{equation}
\item[4)] If the norm of the Kohn-Sham Slater determinant drops below
          the drawn random number $\eta_j $, act with the bath operator(s) on
          the Kohn-Sham orbitals and update the auxiliary orbitals
\begin{align}
||\mbox{Det}\{\phi_j^{\rm aux}(t_k)\}|| \le \eta_k \rightarrow\,\, &
\begin{cases}
 \psi_j^{\rm KS}(t_k) = \hS \psi_j^{\rm KS}(t_k) & \\
 \phi_j^{\rm aux}(t_k) = \psi_j^{\rm KS}(t_k)    &  \nonumber
\end{cases}
\end{align}
\item[5)] Go to step {1)}
\end{itemize}
The piecewise deterministic evolution that is generated by the steps of
this algorithm is illustrated schematically in Fig. \ref{fig:quantum_jump_algorithm}.

Averaging at any given time over an ensemble of
stochastic realizations allows then to obtain mean values of any
physical observable. It is also important to realize that we have a 
full statistical ensemble at hand. This allows to compute {\em distributions} of 
observables, higher order moments, cumulants, etc.
In this sense the ensemble of stochastic realizations generated by the
stochastic Schr\"odinger equation carries more information than
the statistical operator, since the latter is merely a first order moment
and higher order moments and cumulants cannot be computed easily from
the first order moment (if at all).

We also emphasize here, that the interpretation of a single stochastic trajectory
is not meaningful: the stochastic realizations have to
be considered always as an ensemble.
When averages over the stochastic ensemble are performed, the ``convergence'' 
of all observables of interest has to be checked carefully by increasing
the number of realizations of the stochastic process.

Note that without further constraints the action of the bath
operator in step 4) of the algorithm can in principle lead to a
loss of orthogonality. For example, all orbitals of the Slater
determinant could relax to the same orbital shape. The system could
loose in this way its fermionic character. In order to maintain the
fermionic nature of the Kohn-Sham state vector, we have to ensure
that the orbitals of the Kohn-Sham Slater determinant remain
orthogonal. To achieve this, we perform an orthogonalization of
the orbitals after each action of the bath operators. This
orthogonalization can be thought of as being part of the
definition of the action of the operator $\hat S$.

From our numerical experience so far, the waiting-time distribution
seems to follow mainly a single exponential distribution. It would
therefore appear appealing to parametrize this distribution and to
draw the waiting times from the analytical expression of the
parametrization. In this way the propagation of the auxiliary states
could be avoided and a speedup of the propagation by a factor of
two could be gained. However, it is not clear if the waiting times
of the Kohn-Sham system follow always an exponential distribution.
In particular, the shape of the distribution is unknown, when, e.g., ionic
motion is involved, or when the system is subject to strong external
electric or magnetic fields.
Therefore, to be on the safe side, in the present work we always
sample the waiting-time distribution by propagating the auxiliary
system.

It is also worth noting that the average over stochastic realizations of
the ensemble generally converges faster when the system-bath interaction
increases. In the opposite limit the convergence is slow. When
the system-bath interaction is very weak, only a small damping will be
exerted by the $\hS^\dagger\hS$ term in Eq.~(\ref{phieq}), and hence
it takes longer for the norm of the auxiliary wave functions to drop
below their waiting times. This in turn implies that fewer jumps
occur and hence more stochastic realizations are required to converge to
a smooth observable distribution.

\section{Applications}
\label{sec:applications}

\subsection{Finite Systems}

\begin{figure}[!h!t]

\centering
\includegraphics[scale=1,width=0.49\columnwidth]{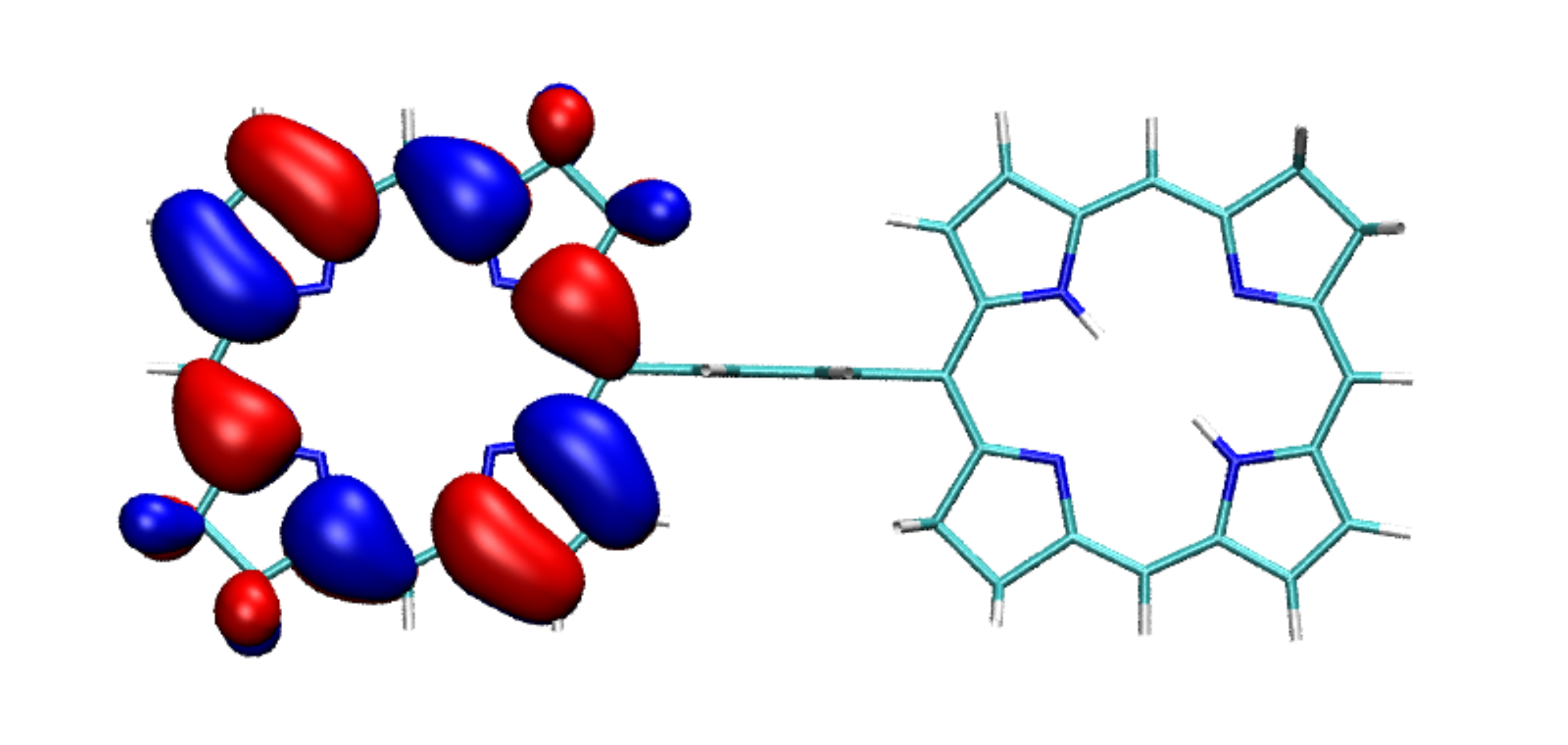}
\includegraphics[scale=1,width=0.49\columnwidth]{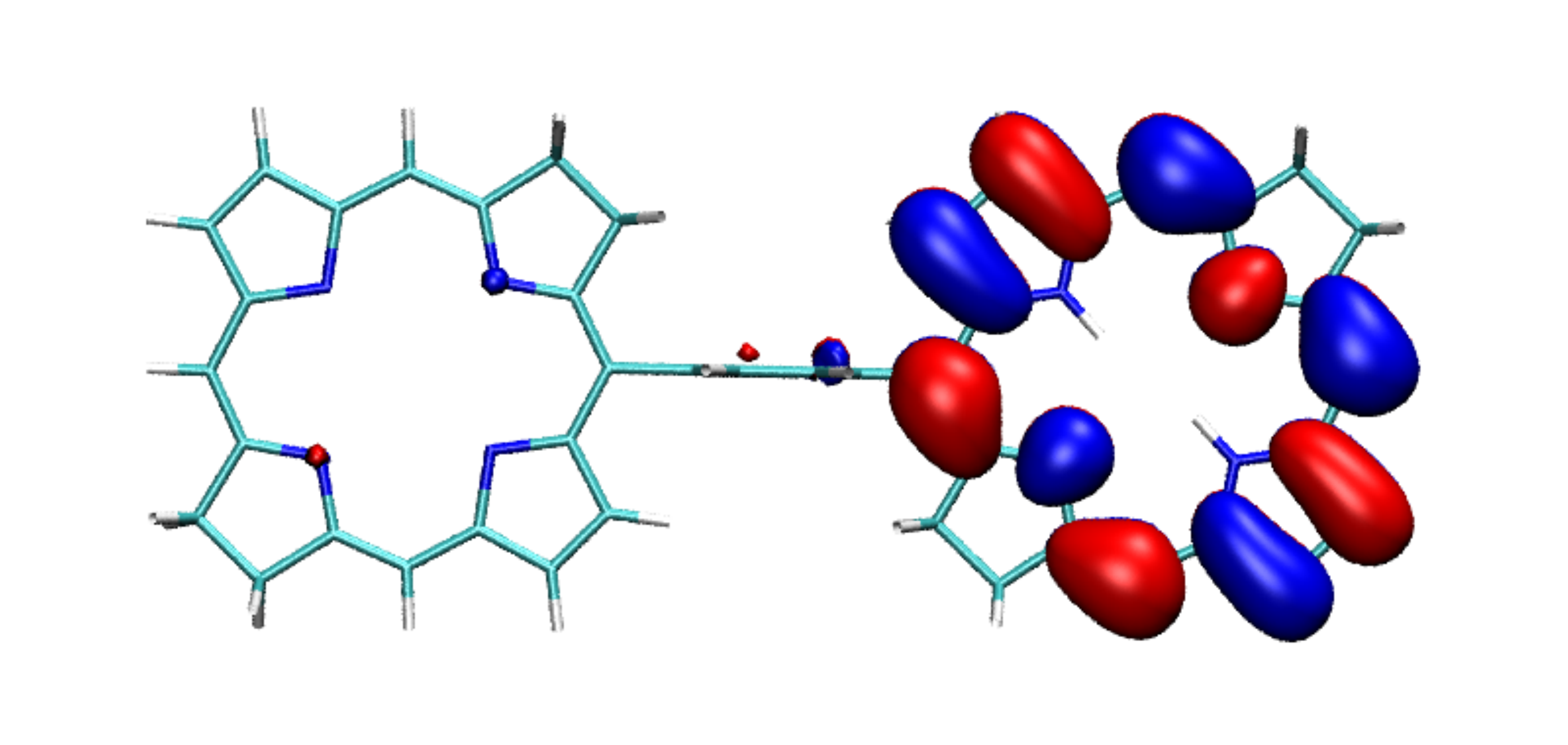}
 \caption{
  Real part of the HOMO (left panel) and LUMO (right panel)
  orbitals of (1,4)-phenylene-linked zincbacteriochlorin-bacteriochlorin.
 \label{fig:bacteriochlorin_homo_lumo}}
\end{figure}

In the last section we have introduced technical details for
the quantum jump algorithm that we use to simulate the stochastic
process associated with the stochastic Kohn-Sham equation, Eq.~(\ref{eq:kssesp}).
In this section we apply the algorithm to molecular systems with and
without clamped ions. As first example we consider a situation of clamped ionic
coordinates. As testcase we investigate a (1,4)-phenylene-linked
zincbacteriochlorin-bacteriochlorin complex. Due to an extra Mg
atom in the left porphyrin ring of the complex the molecule
is not fully symmetric.
As a result, the highest-occupied molecular orbital (HOMO) of this molecule
is located on the left porphyrin ring, whereas the lowest-unoccupied molecular
orbital (LUMO) is located on the right porphyrin ring, cf.
Fig.~\ref{fig:bacteriochlorin_homo_lumo}.
%
%
\begin{figure}[!h!t]
\centering
\includegraphics[scale=1,width=0.9\columnwidth]{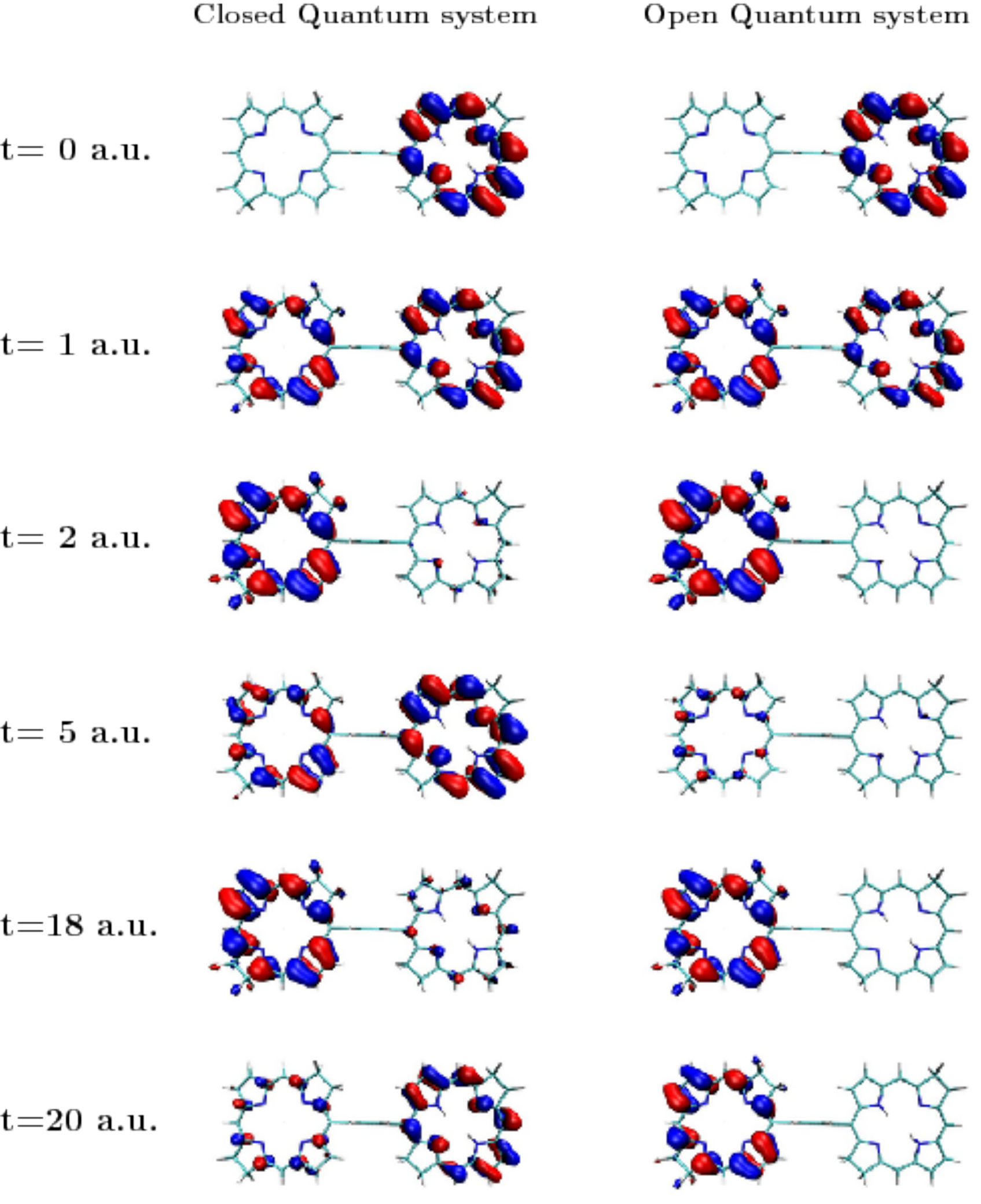}
 \caption{
  Snapshots of the time evolution of the HOMO orbital of
  zincbacteriochlorin-bacteriochlorin with clamped ions.
  The plots display the real part of the orbital.
  In the left column the closed quantum system evolution is
  shown at different points in time and the right column
  displays the evolution of the system with a coupling to
  a thermal bath. A rather fast relaxation rate of $\tau = 1$ a.u.
  has been used for the bath operator in the open quantum
  system case.
 \label{fig:bacteriochlorin}}
\end{figure}
This system has been used as a model to study charge-transfer
excitations in linear response TDDFT \cite{dreuw-2004}.
Here instead we consider open and closed system real-time propagation.
We prepare the zincbacteriochlorin-bacteriochlorin complex in an
entangled initial-state, where the orbital of the HOMO is replaced by
\begin{equation}
\psi_{\HOMO}^{\TDKS}(t=0) = \frac{1}{\sqrt{2}}
\left[ \psi_{\HOMO}^{\GS} + e^{-i \frac{\pi}{2}} \psi_{\LUMO}^{\GS}  \right],
\label{eq:entangled_initial_state}
\end{equation}
where $\psi_{\HOMO}^{\GS}$ and $\psi_{\LUMO}^{\GS}$ denote the
ground-state HOMO and LUMO, respectively.
For all other orbitals the ground-state configuration is used
at the initial time.
Starting from this excited initial Slater determinant the system
is then evolved freely in time without any external fields. For
the bath operators we employ the model of Eq.~(\ref{eq:bath_operators})
introduced in section \ref{sec:system_bath_interaction_model} at zero temperature.
The dynamics of the system is illustrated in Fig.~\ref{fig:bacteriochlorin}
where we plot the real part of the HOMO orbital for different snapshots
in time. The left panel summarizes the closed system evolution
and the right panel an ensemble average over 100 stochastic
realizations in the open system case. Let us first focus on the
closed quantum system case. Due to the entangled
initial state in Eq.~(\ref{eq:entangled_initial_state}), the
time-dependence of the orbital $\psi_{\HOMO}^{\TDKS}(t)$ has mainly
oscillatory phase contributions $\exp(-i\epsilon_{\rm HOMO}t)$ and
$\exp(-i\epsilon_{\rm LUMO}t)$ from the ground-state HOMO and LUMO,
respectively. Only small nonlinearities arise due to the dependence
of the Kohn-Sham Hamiltonian on the time-dependent density. Since the
system is propagated as closed quantum system, the phase oscillations would
continue indefinitely.

On the other hand, the open quantum system evolution in the right
panel of Fig.~\ref{fig:bacteriochlorin}  shows
for $\psi_{\HOMO}^{\TDKS}(t)$ a fast relaxation from the entangled
initial state back to the HOMO which is localized on the left porphyrin
ring.
If we now imagine computing Ehrenfest forces from these orbital
contributions, it is clear that the forces will differ qualitatively
in the closed and open quantum system cases. While in the closed
quantum system case the forces will be
oscillatory, they will show relaxation behavior similar
to the orbitals in the open quantum case. This example
emphasizes that the coupling of electronic degrees of freedom to a
thermal bath yields {\em qualitatively} different forces compared
to standard QMD approaches.

\begin{figure*}[!h!t]
 \includegraphics[width=0.95\columnwidth]{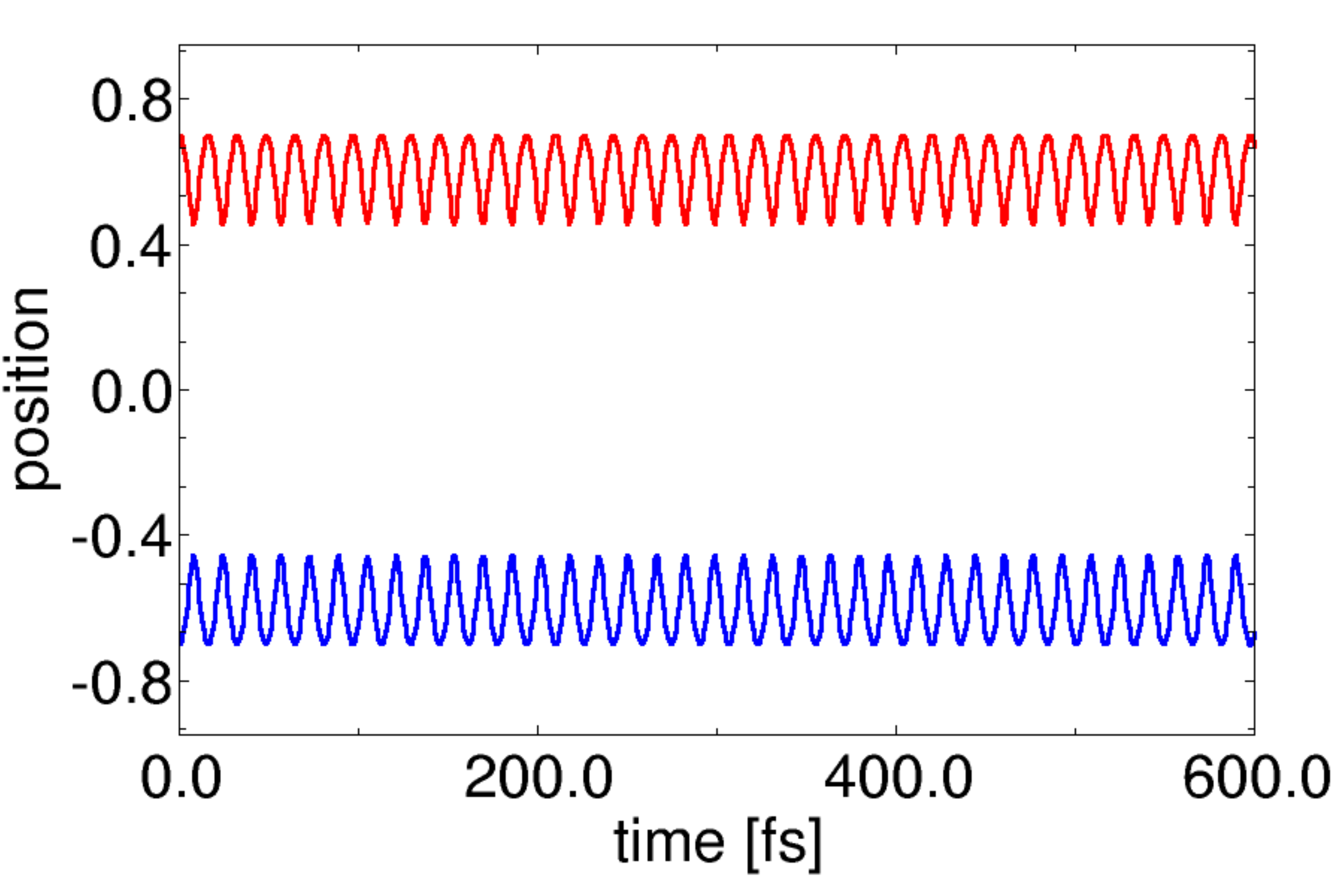}
 \hspace{0.09\columnwidth}
 \includegraphics[width=0.95\columnwidth]{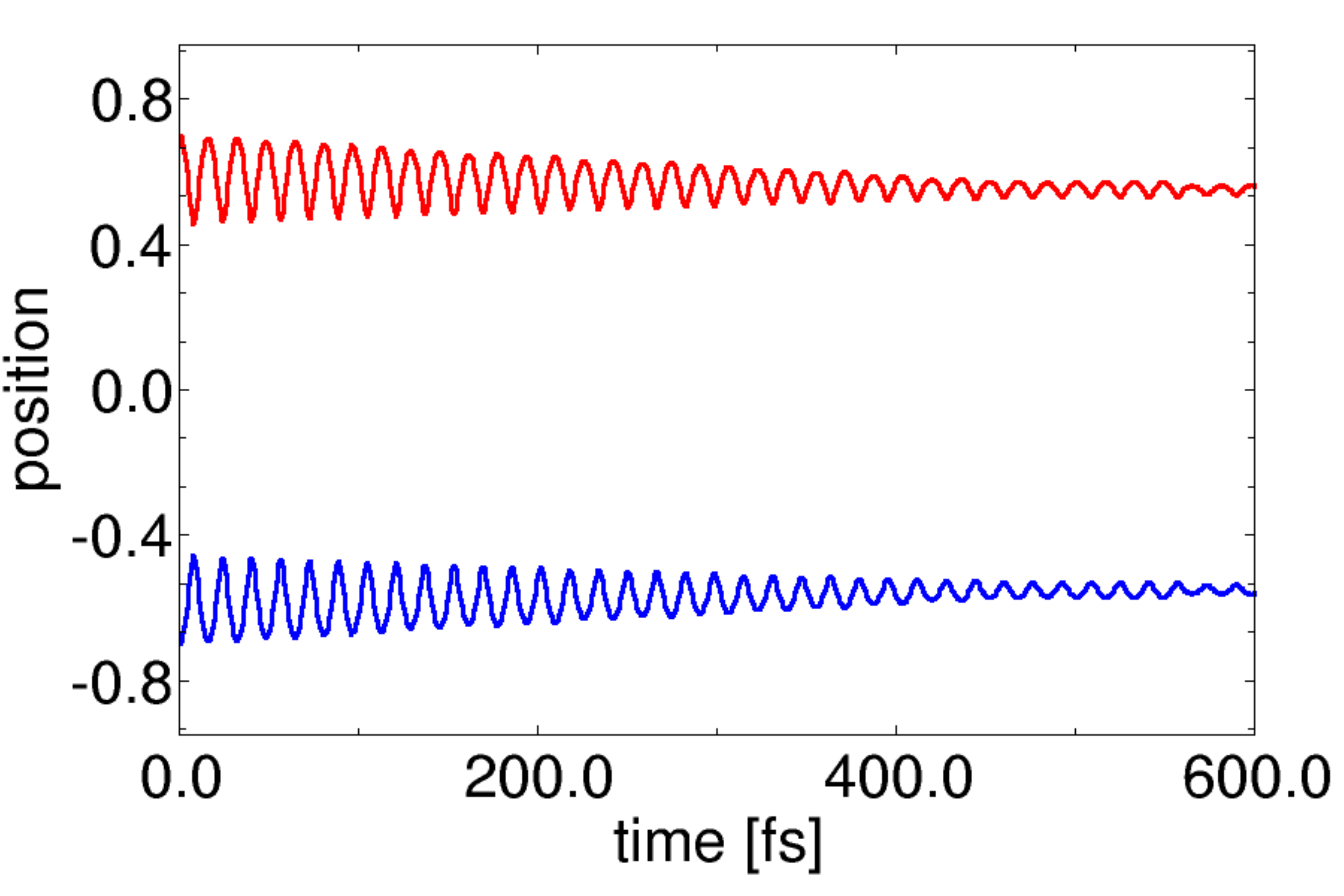}
 \caption{
  Left panel: Here we show the ionic positions of a Neon dimer
  as function of time for a closed quantum system. As initial
  condition we have selected a stretched configuration of the
  dimer which results in an indefinite coherent oscillation of the
  two nuclei.
  Right panel: Using the same initial state we have evolved with
  SQMD a stochastic ensemble of trajectories. Shown is the average
  of the nuclear positions for an ensemble with 100 stochastic
  realizations. As relaxation time for the simulation we have employed
  $\tau = 300$~fs. The ionic velocities follow a Maxwell-Boltzmann
  distribution with a temperature of 290K.
 \label{fig:ion-damping-closed}}
\end{figure*}
This observation motivates our second example, where we consider a stochastic
QMD simulation for a neon dimer. In this case the ions are not
clamped at the equilibrium configuration. Instead we use stretched
initial positions for the ions of the dimer as initial state for the
open and closed system propagation.
If we would treat the ions quantum mechanically, then the bath operators
would also act on the nuclear wavefunctions. However, since we have restricted
ourselves here to the limit of classical ions we replace this action of the
bath operators by modifying the velocities of the ions. At every occasion
when the bath operators act on the electronic wavefunctions we draw new
velocities for the ions from a Maxwell-Boltzmann distribution. This is a
simple approach but can be improved with e.g., recently introduced stochastic
thermostats for the ions \cite{bussi-2007}.

In the left panel of figure
Fig.~\ref{fig:ion-damping-closed} we show the ionic positions of
the dimer as function of time for a standard Ehrenfest TDDFT closed
quantum system evolution. In the right panel we display for the
same initial conditions the open quantum system evolution within SQMD.
For the SQMD simulation we have employed a relaxation time of $\tau = 300$ fs
and an average over 100 stochastic trajectories has been
performed which results in a smooth decay of the nuclear oscillations.

\subsection{Extended Systems}
So far we have considered only finite molecular systems.
However, a large class of applications requires also the
treatment of periodic boundary conditions in one, two or
three dimensions. This includes for example decoherence and
dissipation in nanowires, electronic relaxation on surfaces,
or hot electron thermalization in bulk systems.
For these cases it is desirable to extend our approach to
periodic systems. In this section we briefly discuss the
necessary steps in order to apply SQMD to extended systems.

There are some extra details and conditions that have to be
satisfied in order to treat periodic systems with SQMD.
As a first step we expand the stochastic Kohn-Sham orbitals
of the periodic system of interest in the complete set of
the corresponding ground-state Bloch-orbitals $\varphi_k(\br)$
\begin{equation}
\psi_k(\br, t) = \sum_{k'} d_{k'}(t) \varphi_{k'}(\br).
\end{equation}
This gives rise to stochastic expansion coefficients
$d_{k}(t)$ which are then propagated in time using, e.g., the
quantum jump algorithm that we have presented in
section~\ref{QJA}. Similar to the case of molecules, the fermionic
nature of the electronic subsystem needs to be taken into account
by orthogonalizing the occupied states after each
application of the bath operator (cf. step (4) of the quantum
jump algorithm).

In addition, care has to be taken for the choice of gauge for the
vector and scalar potentials in the Hamiltonian
of the extended system. Here, the same restrictions apply
as in standard closed-system TDDFT simulations. In practice,
we consider only purely time-dependent vector potentials
which retain the periodicity of
the considered system at all times. In the present context we
have to assume in addition that the bath operators retain the
periodicity of the extended system as well. This restricts the
choice of baths represented by local operators that satisfy the
condition
\begin{equation}
\hat S(\br) = \hat S(\br + {\bf R})
\end{equation}
where ${\bf R}$ denotes the usual displacement vector of the unit cell.
This may exclude certain relaxation mechanisms. However,
the importance of these relaxation and dephasing channels
can always be checked by increasing the size of the supercell
that is used in the simulation.

While we do not have fully implemented this scheme yet, we want to
argue about important physical processes that
can be studied with this approach. For instance, one could study
adsorption of molecules on surfaces whose opposite side is set
on a thermal stage that keeps the electron and/or ion
temperatures fixed. Again, this could be accomplished in a
supercell geometry by coupling some ``bulk'' layers away from the
surface with a local operator that
maintains energy equilibrium in that region (an example of such
operator is given in Ref. \cite{dubi-2009}). The rest of the system
is let to follow its own dynamics. If we excite the molecules and/or
surface - e.g., by application of a short electromagnetic field -
electrons and ions can then distribute energy and momentum first in
the layers adjacent to the surface and then relax energy into the
bath, where they would thermalize to the appropriate canonical
distributions. Analogously, we could monitor energy relaxation of
electrons and ions in a bulk exited either thermally or electrically
and kept at a given temperature by a thermal stage.
Important phenomena that are then accessible would be, e.g., phase
transitions driven by dissipative effects.

\section{Conclusions}\label{sec:conclusions}
In summary, we have presented a detailed account of stochastic
quantum molecular dynamics. The approach is based on a stochastic
Schr\"odinger equation, which may or may not describe Markovian
dynamics - although we have focused the discussion to the Markovian
case. Our approach allows us to describe the dynamics of electrons
and ions coupled to one or many external environments. For simplicity
we have restricted our examples to the situation of classical
ions, but the approach is, in principle, valid also for quantum ions.
Although we have not reported any actual implementation of SQMD for
periodic systems, we have outlined the theory behind its extension
to this important case. Work along these lines is in progress and
will be reported elsewhere \cite{appel-2011}.

This approach is thus amenable to studying many interesting phenomena
related to energy relaxation and dephasing of the electronic subsystem
in the presence of ionic dynamics, such as local ionic and electronic
heating in laser fields, relaxation processes in photochemistry, etc.,
a feature that is lacking in any ``standard'' molecular dynamics approach.


We acknowledge support from DOE
under grant DE-FG02-05ER46204 and Lockheed Martin.


\end{document}